%Paper: gr-qc/9302023
%From: STEP8030@bureau.ucc.ie
%Date: Fri, 19 Feb 1993 12:36 GMT

% CUP macro
% FONT DEFINITIONS
% ----------------------------------------------------------------

\font\sixrm=cmr6
\font\sixi=cmmi6
\font\sixsy=cmsy6

\font\sevenrm=cmr7
\font\seveni=cmmi7
\font\sevensy=cmsy7

\font\twelverm=cmr12
\font\twelvei=cmmi12
\font\twelvesy=cmsy10 at 12pt
\font\twelveit=cmti12
\font\twelvesl=cmsl12
\font\twelvebf=cmbx12
\font\twelvett=cmtt12

\def\twelvepoint{%
\def\rm{\fam0\twelverm}%
\def\it{\fam\itfam\twelveit}%
\def\sl{\fam\slfam\twelvesl}%
\def\bf{\fam\bffam\twelvebf}%
\def\tt{\fam\ttfam\twelvett}%
\def\cal{\twelvesy}%
 \textfont0=\twelverm
  \scriptfont0=\sevenrm
  \scriptscriptfont0=\sixrm
 \textfont1=\twelvei
  \scriptfont1=\seveni
  \scriptscriptfont1=\sixi
 \textfont2=\twelvesy
  \scriptfont2=\sevensy
  \scriptscriptfont2=\sixsy
 \textfont3=\tenex
  \scriptfont3=\tenex
  \scriptscriptfont3=\tenex
 \textfont\itfam=\twelveit
 \textfont\slfam=\twelvesl
 \textfont\bffam=\twelvebf
 \textfont\ttfam=\twelvett
 \baselineskip=15pt
}

\font\sixteenrm=cmr17 at 16pt
\font\twentyrm=cmr17 at 20pt

% PAGE MAKEUP PARAMETERS
% ----------------------------------------------------------------

\hsize     = 152mm
\vsize     = 215mm
\topskip   =  15pt
\parskip   =   0pt
\parindent =   0pt

% NEW COMMANDS
% ----------------------------------------------------------------
\newskip\one
\one=15pt
\def\One{\vskip-\lastskip\vskip\one}

\newcount\LastMac
  % null element
\def\Skipe{1}  % SkipToFirstLine
\def\Txe{2}    % text
\def\Hae{3}    % heading A
\def\Hbe{4}    % heading B
    % heading C

\def\SkipToFirstLine{% move to start of text proper
 \LastMac=\Skipe
 \dimen255=150pt
 \advance\dimen255 by -\pagetotal
 \vskip\dimen255
}

\def\Raggedright{%
 \rightskip=0pt plus \hsize
 \spaceskip=.3333em
 \xspaceskip=.5em
}

\def\Fullout{% justify all lines
 \rightskip=0pt
 \spaceskip=0pt
 \xspaceskip=0pt
}

% DESIGN ELEMENTS
% ----------------------------------------------------------------

\def\ct#1\par{% chapter title
 \One
 \Raggedright
 \twentyrm\baselineskip=24pt
 #1
}

\def\ca#1\par{% chapter author
 \One
 \Raggedright
 \sixteenrm\baselineskip=18pt
 \uppercase{#1}
}

\def\aa#1\par{% author affiliation
 \One
 \Raggedright
 \twelverm\baselineskip=15pt
 #1
}

\def\ha#1\par{% heading A
 \ifnum\LastMac=\Skipe \else \One\fi
 \LastMac=\Hae
 \Raggedright
 \twelvebf\baselineskip=15pt
 \uppercase{#1}
}

\def\hb#1\par{% heading B
 \LastMac=\Hbe
 \One
 \Raggedright
 \twelvebf\baselineskip=15pt
 #1
}

\def\tx{% text
 \ifnum\LastMac=\Hae \else
  \ifnum\LastMac=\Hbe \else
   \ifnum\LastMac=\Skipe \else \One
   \fi
  \fi
 \fi
 \LastMac=\Txe
 \Fullout
 \twelvepoint\rm
}

\def\bb{% bibliography/references
 \One
 \Fullout
 \twelvepoint\rm
  \hangindent=1pc
  \hangafter=1
}

%%%\output{\OutputPage}

\ct Brill Waves \par
\ca Niall \'O Murchadha \par
\aa Physics Department, University College, Cork, Ireland
\par

\SkipToFirstLine
\hb Abstract \par

\tx Brill waves are the simplest (non-trivial) solutions
to the vacuum constraints of general relativity. They
are also rich enough in structure to allow us believe
that they capture, at least in part, the generic
properties of solutions of the Einstein equations. As
such, they deserve the closest attention. This article
 illustrates this
point by showing how Brill waves can be used to
investigate the structure of conformal superspace.

\ha 1 Introduction \par

\tx From time to time I amuse myself by
mentally assembling a list of articles I would like to have
written. The candidates for this list have to satisfy a
number of criteria. Naturally, they have to be both
important and interesting to me. Equally, they have to
contain results that I can convince myself, however
unreasonably, that I could have obtained. Every time I
make my list I am struck again by the number of articles
by Dieter Brill appearing on it.  At first glance, this
is explained by the large overlap between our interests.
In reality, however, the explanation is to be found by
considering the kind of article that Dieter has written
over the years and the way in which he manages to convey
major insights in a deceptively simple fashion.
In this work I wish to return to an article that has a
permanent place on my list, the famous Brill waves, and
show how some of the the earliest work that Dieter did in
general relativity continues to offer valuable insight a
third of a century later.

\ha 2 Brill \par

\tx In his thesis (Brill, 1959) Dieter Brill considered
axisymmetric, moment-of-time-symmetry, vacuum initial
data for the Einstein equations.The starting  point is an
axially symmetric three-metric of the following form
$$
g = e^{2Aq}(d\rho^2 + dz^2) + \rho^2d\theta^2, \eqno(1)
$$
where $A$ is a constant and $q$ is an (almost) arbitrary
function of $\rho$ and $z$. We only require that it satisfy
$q = q_{,\rho} = 0$ along the z-axis, that it decay fairly
rapidly at infinity (faster than 1/r) and that it be
reasonably differentiable.

\tx This metric $g$ is to be conformally transformed to a
metric $\bar{g} = \phi^4g$ so that the metric $\bar{g}$
has vanishing scalar curvature, so as to satisfy the
moment-of-time-symmetry initial value constraint of the
Einstein equations. This is  equivalent to seeking a
positive solution $\phi$ to $$ 8\nabla^2_g\phi - R\phi =
0,\ \ \phi\ >\ 0,\ \ \phi \ \rightarrow \ 1 \ \ at\
\infty, \eqno(2) $$
where $R$ is the scalar curvature of $g$. It is easy to
calculate
$$
R = -2A e^{-2Aq}(q_{,\rho\rho} + q_{,zz}). \eqno(3)
$$
It is important to notice that $\sqrt{g} = \rho e^{2Aq}$
so that
$$\eqalignno{
\int R dv &= \int \sqrt{g} R d^3x = -4A\pi \int \rho
(q_{,\rho\rho} + q_{,zz}) d\rho dz \cr
&= -4A\pi \int
[(\rho q_{,\rho} - q)_{,\rho} + (\rho q_{,z})_{,z}] d\rho
dz \cr
&= 0. &(4)}
$$
To obtain the last line we have to use both the
regularity along the axis and the asymptotic falloff of
$q$. This is a remarkably powerful result, in that it
gave Brill the first positivity of energy result in
General Relativity.

\tx Let us assume that we obtain a regular solution to
(2) (we will return to this issue in Section 3). The
total energy, the ADM mass, is contained in the 1/r part
of the physical metric $\bar{g}$. Since the base-metric
$g$ falls off faster than 1/r the mass must be contained
in the 1/r part of the conformal factor $\phi$. More
precisely we get
$$
\phi\ \ \rightarrow \ \ 1 + {M \over 2r}. \eqno(5)
$$
Hence
$$\eqalignno{
2\pi M &= -\oint_{\infty}\nabla^a\phi dS_a =
-\oint_{\infty}{\nabla^a\phi \over \phi}dS_a &(6)\cr
&= \int[-{\nabla^2\phi \over \phi} + {(\nabla \phi)^2
\over \phi^2}]dv = \int [-{R \over 8} +{(\nabla \phi)^2
\over \phi^2}]dv &(7)\cr
&= \int{(\nabla \phi)^2 \over \phi^2}dv > 0. &(8)\cr}
$$
Notice that we use (4) in going from (7) to (8). The key
(as yet) unresolved question is: When can we solve (2)?

\ha 3 Cantor and Brill \par

\tx The next issue to be considered is what choices of
$q$ (or $Aq$) allow a regular solution to (2). This
question was first seriously discussed by Cantor and
Brill (1981). They derived the following
\proclaim Theorem I (Cantor and Brill). There is a
$\bar{g}$ conformally equivalent to a given metric $g$
such that $R(\bar{g})$ = 0 if and only if
$$
\int[8(\nabla f)^2 + R(g)f^2]dv > 0, \eqno(9)
$$
for every $f$ of compact support with $f$ not identically
0.\par

\tx Cantor and Brill further show that if $R(g)$ is small
(in a precise sense) then the Sobolev inequality may be
used to guarantee inequality (9). The Sobolev inequality
states that for any asymptotically flat Riemannian
three-manifold there exists a positive constant $S(g)$
such that
 $$
 \int(\nabla f)^2 dv > S(g)[\int f^6 dv]^{1
\over 3}, \eqno(10)
$$
for any $f$ of compact support. Let us now use the
H\"older inequality on the second term in (9) to give
$$
\int R(g) f^2 dv < [\int |R(g)|^{3 \over 2} dv]^{2 \over
3} [\int f^6 dv]^{1 \over 3}. \eqno(11)
$$
Combining (10) and (11) shows that if
$$
[\int |R(g)|^{3 \over 2} dv]^{2 \over3} < 8S(g) \eqno(12)
$$
then expression (9) must be positive for any $f$ and a
regular solution to (2) exists.

\tx The Brill waves supply an obvious application of this
result. Choose a metric of the form (1) with a fixed $q$
but allow $A$ to vary. It is clear from the expression
(3) for the scalar curvature that we can always find a
small enough $A$ (which can be either positive or
negative) to guarantee that (12) holds.

\tx The next interesting property of the Brill waves is
that, for a fixed $q$, one can always find a large enough
value of $A$ ( both positive and negative) such that
inequality (9) cannot hold. We know that the scalar
curvature integrates to zero. This means that it must
have positive and negative regions. It is clear that we
can always find an axisymmetric function $f_+$ which has
support only on the positive regions of the scalar
curvature and another function $f_-$ which has support
only on the negative regions. This choice can be made
independent of the value of $A$. We have that
$$
\int(\nabla f_+)^2 dv = 2\pi \int[(f_{+,\rho})^2 +
(f_{+,z})^2] \rho d\rho dz \eqno(13)
$$
entirely independent of $A$. We also have
$$
\int R(g) f_+^2 dv = 4 \pi A \int[- (q_{,\rho\rho} +
q_{,zz})]f_+^2 \rho d \rho dz \eqno(14)
$$
where the integral on the right hand side of (14) is
positive (from the choice of $f_+$) and independent of
$A$. Therefore, with this choice of test-function there
exists a number $|A_-|$ given by
$$
|A_-| = {\int[(f_{+,\rho})^2 +
(f_{+,z})^2] \rho d\rho dz \over 2\int[- (q_{,\rho\rho} +
q_{,zz})]f_+^2 \rho d \rho dz} \eqno(15)
$$
such that if $A$ is large and negative, i.e., $A <
-|A_-|$, inequality (9) cannot hold. A similar bound,
using $f_-$, can be derived showing that there exists
an $|A_+|$ such that if $A > |A_+|$ (9) again breaks down.

\tx This is not just a mathematical game, conformal
transformations are the standard way of constructing
solutions to the initial value constraints of General
Relativity (\'O Murchadha and York, 1973) and for an
interesting set of such data (`maximal' initial data) a
necessary and sufficient condition is that the metric be
conformally transformable to one with zero scalar
curvature. What we are doing, therefore, is trying to map
out conformal superspace $\tilde{S}$; more precisely, we
are trying to delineate the axially symmetric subspace of
conformal superspace (a so-called `midisuperspace'). The
functions $q(\rho, z)$ can be regarded as defining
`directions' in conformal superspace with $A$ acting as a
`distance' along each given ray. The `point' $A = 0$ is
flat space; there is an open interval around the origin
on each ray which belongs to conformal superspace and
each ray (after a finite `distance') emerges from
$\tilde{S}$.

\tx We can even prove more; we can show that each ray
passes only once out of $\tilde{S}$. Given $q$, let us
assume that we have passed out of  $\tilde{S}$, in other
words we have an $A_0$ (assumed positive) such that
inequality (9) does not hold. This means that there
exists a function $f_m$ such that
$$
\int[8(\nabla f_m)^2 + R(g)f_m^2]dv \leq 0. \eqno(16)
$$
In other words,
$$
4\pi \int(4[(f_{m,\rho})^2 +
(f_{m,z})^2] - A_0 [ (q_{,\rho\rho} +
q_{,zz})]f_m^2)\rho d \rho dz \leq 0.\eqno(17)
$$
The second term must be negative to counter the
positive first term. It is clear that if one increases $A$
while holding $f_m$
 and $q$ fixed the inequality must worsen. Therefore,
once a ray passes outside  $\tilde{S}$, it stays outside.
 In other words, (the axially symmetric part of)
$\tilde{S}$ seems to be simply connected with convex
boundary.

 \ha 4 Beig and \'O Murchadha \par

\tx The next question that needs be asked is what
happens as one moves along such a ray in $\tilde{S}$ and
approaches the critical value of $A$. This question has
been addressed is a number of recent articles (\'O
Murchadha, 1987, 1989; Beig and \'O Murchadha, 1991).

\tx There exists an object analogous to the Sobolev
constant, $S(g)$, defined in equation (10), the conformal
Sobolev or the Yamabe constant. The Yamabe constant is
defined by
$$
Y(g) = inf{\int[8(\nabla f)^2 + R(g)f^2]dv \over
8(\int f^6 dv)^{1 \over 3}}, \eqno(18)
$$
where the infimum is taken over all smooth functions of
compact support. This object has a number of interesting
properties. It is a conformal invariant; it achieves its
maximum value [equalling $3({\pi^2 \over 4})^{2/3}$] at
(conformally) flat space (Schoen, 1984). Flat space is its
only extremum.

\tx One reason for introducing this concept here is that
it allows a different formulation of Theorem I.
\proclaim Theorem Ia.  There is an asymptotically flat
metric $\bar{g}$ conformally equivalent to a given metric
$g$ such that $R(\bar{g})$ = 0 if and only if the Yamabe
constant $Y(g)$ of $g$ is positive.

\tx One can immediately see this by comparing
expressions (9) and (18).

\tx The function that minimizes
the Yamabe functional satisfies the non-linear equation
$$
8\nabla^2_g\theta - R\theta = -8\lambda\theta^5,\ \
\theta \ \rightarrow \ 0 \ \ at\ \infty, \eqno(19)
$$
where $\lambda$ is a constant that is proportional to
the Yamabe constant. It equals the Yamabe constant if
$\theta$ is normalized via
$$
\int\theta^6dv = 1.
$$
I will use this normalization from now on. It can be shown
(Schoen, 1984) that $\theta$ exists, is everywhere
non-zero, and falls off at infinity like 1/r.

\tx Let us now combine $\phi$, the solution of eqn.(2),
and $\theta$, the solution of eqn.(18), using the Green
identity
$$\eqalignno{
-8\oint_{\infty}\nabla^a\theta dS_a &=
8\oint_{\infty}(\theta\nabla^a\phi -
\phi\nabla^a\theta)dS_a & (20) \cr
&= 8\int(\theta\nabla^2\phi - \phi\nabla^2\theta) dv
& (21) \cr
&= \int[\theta R\phi - \phi(R\theta - Y\theta^5)]dv
& (22) \cr
&= Y\int \theta^5 \phi dv. & (23) \cr}
$$
As $Y$ approaches zero, both $\theta$ itself, and the
surface integral of $\theta$ remain well-behaved.
 In particular, the 1/r part of $\theta$ does
not vanish in the limit. Therefore the surface
integral remains bounded away from zero. This
means that the volume integral in (23) must blow
up like 1/Y. Since $\theta$ remains finite, $\phi$ must
become unboundedly large. In other words, $\phi$ must blow
up like 1/Y.

\tx It is not enough that $\phi$ blow up like a delta
function at one point, it must become large on an
extended region so that the integral $\int \theta^5 \phi
dv$ blows up like 1/Y. However, we know that $\phi$
is not a random object, it satisfies a differential
equation which forces a broad blow-up on $\phi$. We can
make this more precise by returning to eqn.(2).  Using the
fact that $\sqrt{g} = \rho e^{2Aq}$, we find that $\phi$
must satisfy
 $$
4\nabla^2_f\phi + A(q_{,\rho\rho} + q_{,zz})\phi = 0,\ \
\phi\ >\ 0,\ \ \phi \ \rightarrow \ 1 \ \ at\ \infty,
\eqno(24)
 $$
where $\nabla^2_f$ is the flat-space laplacian.

\tx Let us consider the situation
 where $q$ has compact support
on some ball $B$. The scalar curvature, $R$, has the
same compact support. Outside this region the manifold is
flat and $\phi$ satisfies $\nabla^2_f\phi = 0$. Hence the
min-max principle tells us that the maximum of $\phi$ must
be achieved inside the ball $B$. Further, since $\phi$
is positive and satisfies a linear elliptic equation on
the compact set $B$ we can use the Harnack inequality
(Gilbarg and Trudinger, 1983). This means that there
exists a constant $C_B$ (which we can choose independent
of $A$) such that
$$
 \max \; \phi = \max_B\phi \leq C_B\;\min_B \phi \leq
C_B\;\min_{\delta B}\phi. \eqno(25)
$$
In other words, as the maximum of $\phi$ increases
inside in $B$, $\phi$ everywhere in $B$ gets dragged
up with it. In particular, $\phi$ on the boundary of
$B$ becomes large.

\tx As $A$ gets large, as $Y$ gets
small, we reach a point where $\max\; \phi \geq C_B$ (the
Harnack constant). From that point on we have that
 $min_{\delta B}\phi \geq 1$. Outside $B$ $\phi$
satisfies $\nabla^2\phi = 0$. This means that the
minimum and maximum of $\phi$ occur on the boundaries of
the set. Since $\phi \geq 1$ on $\delta B$ and $\phi$
= 1 at infinity, the minimum must be at infinity.
Returning to (5), $\phi\  \rightarrow \  1 +
{M \over 2r}$, we can conclude that the constant $M$
must be positive. This is a positive energy proof which
is not as strong as the original Brill proof, it only
holds in the strong-field region, far away from flat
space, as the Yamabe constant becomes small. However, it
has the advantage of working for a much larger class of
metrics.

\tx By analogy with electrostatics we can regard $\phi$
(or rather $\xi = \phi - 1$) as the potential  on
and outside a (nonconducting) shell ($\delta B$). The
potential on the shell is given and one solves for the
potential outside. The ADM mass corresponds to the total
charge that one has to distribute on the shell to give
the specified potential distribution. If the potential
is everywhere positive on the shell, the charge must
be positive.

\tx If the shell were conducting, the potential on it
would be constant. Now we can talk about the
capacitance $C$, the constant ratio between the `charge'
$M$ and the potential on the surface. The capacitance
is a kind of Harnack constant because we have
$$
C\; \min _{\delta B}\xi \leq M. \eqno (26)
$$
$$
C\; \max _{\delta B}\xi \geq M. \eqno(27)
$$
These are easy to derive. Replacing a non-constant
potential on $\delta B$ by its minimum (maximum) value
must decrease (increase) the charge while the decreased
(increased) charge equals
 the capacitance multiplied by the potential. Since the
capacitance depends only on the size and shape of the
surface, not on the charge, we immediately see that as
$\min _{\delta B} \xi$ (or $\min _{\delta B}\phi$)
increases , so will the ADM mass. If we incorporate
(25), (26) and (27) together we can see that the mass
grows linearly with $ \phi$. More precisely, we have
$${C \over C_B}\max \phi \leq M \leq C\max \phi,
\eqno(28) $$
$$C \min _B \phi \leq M \leq C.C_B \min_B \phi.
\eqno(29)$$

\tx I can extract more information from (23). We have
$$
Y \int \theta^5 \phi dv \geq Y \int_B \theta \phi dv
\geq ( Y \min_B \phi) \int \theta^5 dv. \eqno(30)
$$
The Harnack inequality (25) tells us that if $Y
\max\;\phi$ diverges as $Y$ goes to zero, so also will $Y
\min_B \phi$. Eqn. (28) now tells us that $Y \int \theta^5
\phi  dv$ must become unboundedly large, which contradicts
(23). Hence $Y \max\;\phi$ remains bounded away from both
zero and infinity as $Y$ goes to zero. In turn, this
means that both $Y \min_{\delta B} \phi$ and
$Y \max_{\delta B} \phi$ remain finite and bounded away
from zero as $Y$ goes to zero. Finally, this means that
the ADM mass blows up like 1/Y as $Y$ goes to zero.

\tx Not only does the monopole blow up like 1/Y, all
the other multipoles behave in a similar fashion. This
allows us to show that a minimal two-surface must appear
in the conformally transformed space if M becomes
large enough. Let us consider a surface which has
coordinate radius r in the background (flat) space.
The area of this surface in the physical space is A =
$4\pi r^2 \phi^4$. $ A_{,r}$  negative is
equivalent to $\phi + 2r \phi_{,r} < 0$. The leading
terms in $\phi$ are
$$
\phi = 1 + {M \over 2r} + Q{r^2 - 3z^2 \over r^5} +
\dots ,\eqno(31)
$$
where Q is the quadrupole moment. Now we get
$$
\phi + 2r{d\phi \over dr} = 1 - {M \over 2r} - 5Q{r^2 -
3z^2 \over r^5} + \dots \eqno(32)
$$
We need to find out if the right hand side of (32) can be
negative. We know that $M$ will become large and
positive but we have no control over $Q$ except that
the ratio Q/M will remain bounded. A similar result
holds for all the other multipole moments. We can bound the
third term by $10|Q|/r^3$. Let us evaluate (32) at r =
M/4. This gives   $$ \phi + 2r{d\phi \over dr} \leq 1 - 2
+ {640|Q| \over M^3} + \dots \eqno(33)
$$
It is clear that as M becomes large (as $Y$ becomes
small) the third term (and all the higher order terms)
becomes small. This implies that the area of the
surface in question reduces when it is pushed outwards.
(Technically we should evaluate the change in area
along the outer normal rather than along the radial
vector, however, the difference between the two becomes
small as M and $r$ become large.) This means that it is an
outer trapped surface in the terminology of Penrose
(1965). There is a minimal area surface outside all the
trapped surfaces; this is the apparent horizon.

\tx As M increases, the apparent horizon approximates more
and more the spherical surface with radius $r = M/2$. The
conformal factor on this surface equals 2, so the
proper area of the apparent horizon approximates
$16\pi M^2$. The location of the apparent horizon moves
outward (relative to the flat background) and its
area increases as M goes to infinity.

\tx When M reaches infinity, when the horizon reaches
infinity, when the Yamabe constant goes to zero, as A
approaches the critical value, something catastrophic
happens. The solution to (2) blows up and we can no longer
construct an asymptotically flat manifold with zero scalar
curvature. On the other hand we have a solution to
(19) with $\lambda$ = 0. This is a conformal factor
that transforms the given base manifold into a regular
compact manifold with zero scalar curvature. Thus one
can construct a closed, vacuum cosmology at moment of
time symmetry. If we increase A beyond the critical value,
we recover a finite solution to (2), but at the price of
allowing $\phi < 0$ in a region. This is John Wheeler's
`Bag of Gold'.

\ha 5 Wheeler \par

\tx  Let me move back in time. The earliest application
I know of the Brill Wave solution was by John Wheeler
in 1964. He considered eqn.(24) and brought his
understanding of the Shr\"odinger equation to bear on
it. He realised that it was just a scattering problem
off a localized potential. The `potential' ($R$)
averages to zero, but Wheeler realized that the
negative parts of the potential were much more
important than the positive parts. He threw away the
positive parts and considered only a negative
potential. He even further simplified the problem and
considered the problem of scattering off a negative
spherical square well potential. He wrote down an
analytic solution to this problem.

\tx Wheeler considered a spherically symmetric square
well potential in flat space, i.e.,
$$ u = -B,\ \ \ r \leq a$$
$$ u = 0,\ \ \ \ r > a. \eqno(34) $$
It is easy to solve

$$\nabla^2\psi - u\psi = 0 \ \ \ \ \psi \rightarrow 1
\hbox{ at } \infty. \eqno(35) $$
The solution is
$$\eqalignno{
\psi &= {1 \over \cos B^{1/2}a}{\sin B^{1/2}r \over
 B^{1/2}r} \ \ \ r \leq a \cr
&= 1 + {a \over r} \bigg( {\tan B^{1/2}a \over  B^{1/2}a}
- 1 \bigg) \ \ \ \ r > a. &(36) \cr}$$
So long as $ B^{1/2}a < \pi/2$, $\psi$ in (36) is well
behaved. As $ B^{1/2}a$ approaches $\pi/2$, both $\psi$
itself, and the coefficient of the 1/r part (the ADM
mass equivalent) become unboundedly large. Further, the
surface on which $\psi = 2$ (the analogue of the apparent
horizon) moves out towards infinity.

\tx At the critical point ($ B^{1/2}a = \pi/2$), $\psi$
ceases to exist. We now obtain a solution to a related
equation
$$ \nabla^2\bar{\psi} - u\bar{\psi} = 0,\ \ \ \
\bar{\psi} \rightarrow 0 \hbox{ at } \infty. \eqno(37)$$
$$\eqalignno{
\bar{\psi} & = {\sin B^{1/2}r \over
 B^{1/2}r} \ \ \ r \leq a \cr
 & = {1 \over B^{1/2}r}\ \ \ \ \ r > a.&(38) \cr}$$
This is the equivalent of the Yamabe constant going to
zero.

\tx In retrospect, I am amazed how accurately this
highly simplified model captures the behaviour of the
actual situation. When the well is shallow, we have a
regular scattering solution and the energy of the
lowest eigenstate is positive. As the well is
deepened the energy of the lowest state moves down
towards zero. As this happens, we get the phenomenon
of resonant scattering, the scattering solution
grows bigger and bigger. When the well is just deep
enough to give us a zero energy bound state, the
scattering state blows up and ceases to exist. If
the well is made even deeper, the scattering
solution reappears but now it is no longer
everywhere positive, it has a node. This pattern
repeats itself as the energy of the next lowest
energy state approaches zero, and so on.

\tx The 1/r part of the scattering
solution also blows up as one approaches the
critical point. If one approaches the
critical point from the other side, where the scattering
solution has a node, the coefficient of the 1/r term
approaches $-\infty$. This now allows one to give a fairly
nonsensical answer to a fairly unreasonable
question: What is the energy of a closed universe?
My answer is zero, the average of $+\infty$ and
$-\infty$!

\tx We now have a fairly precise understanding of
what happens as we pump up one of the Brill waves;
the Yamabe constant approaches zero for some finite
value of the parameter $A$, the mass approaches
$+\infty$ and an apparent horizon appears which
moves out towards infinity. The horizon, as it moves
outwards, becomes more and more spherical. The breakdown
of the system coincides with the horizon reaching
spacelike infinity. At the critical point we get data
which gives us a smooth compact manifold, without
boundary, with zero scalar curvature.

\tx I would like to point out here
that we have less understanding of what happens as we
further increase the factor $A$. Our only real guide is
the Wheeler model. If this is valid, the mass should drop
to $-\infty$ just on the other side (we would have initial
data with a naked singularity), the mass would build up
again until it became positive and then blow up to
$+\infty$ as the eigenvalue of the first excited state
approached zero, and so on. It would be nice to prove
that this is (or is not) the correct picture.

\tx The analysis given in Section 4 only deals with the
situation when the Yamabe constant is close to zero.
Does it hold in general? In particular, if one takes a
Brill wave and increase the constant does the mass
increase monotonically, does the Yamabe constant decrease
monotonically? These are questions that are yet to be
answered.

\tx An even more fundamental question: What happens as
one approaches the boundary of conformal superspace
along a different direction, i.e., not by holding $q$
fixed and increasing $A$, but by changing $q$? I am
unable to answer this question definitively, all I can
say is that it is currently being investigated by
numerical techniques.

\ha 6 Abrahams, Heiderich, Shapiro and Teukolsky \par

\tx Brill waves have been used by the
numerical relativity community from its earliest days
(e.g. Eppley, 1977) and this use has continued right up to
the present.  The standard approach (pioneered by
Eppley) is to choose a metric of the form (1), with $q$
chosen to be some analytic function. Equation (2) is then
solved numerically to find the conformal factor $\phi$.
Finally, the `physical' metric $\bar{g} = \phi^4 g$ is
constructed. Eppley showed numerically that the overall
structure described in the previous sections
holds true. With a small scale factor $A$, equation (2)
can be solved. As $A$ increases, the conformal factor
increases in the middle. Eventually a minimal surface (an
apparent horizon) appears in the physical space.
Finally, at some finite value of $A$, the solution to
(2) blows up.

\tx Recently, Abrahams, Heiderich, Shapiro and Teukolsky
(1992) showed that this picture of conformal superspace
was grossly oversimplified. They chose as $q$ the function
$$
q = \rho^2 \exp [-({\rho^2 \over \lambda^2_{\rho}} + {z^2
\over \lambda^2_z})]. \eqno(39)
$$
Now, instead of holding $q$ fixed in the traditional
fashion, they varied $q$ by reducing the `characteristic
wavelength' $\lambda_{\rho}$ slowly while holding
$\lambda_z$ fixed. Further, they kept changing the scale
factor $A$ so that the ADM mass of the physical metric
remained at the constant value M = 1. They showed that
the scale factor $A$ grew rapidly, $A \propto
\lambda^{-2}_{\rho}$ as $\lambda_{\rho} \rightarrow
 0$. It is clear
that for all values of $\lambda_{\rho} \neq 0$,the
base metric and the physical metric exist and are regular,
asymptotically flat, axially symmetric Riemannian metrics
with positive Yamabe constant. However, both the base
metric and the physical metric
 become singular when $\lambda_{\rho} = 0$. Most
importantly, nowhere along the sequence do apparent
horizons appear in the physical metric.

\tx This analysis raises major questions about the simple
description of conformal superspace obtained in Section
4 by holding $q$ fixed and changing $A$. AHST are also
probing the boundary of conformal superspace and seem to
have discovered that the boundary, in addition to smooth
metrics with $Y$ = 0, also contains singular metrics.

\tx In the spirit of AHST, let me consider the following
sequence of metrics
$$
g_{\lambda} =  e^{2Cq_{\lambda}}(d\rho^2 + dz^2) +
\rho^2d\theta^2, \eqno(40)
 $$
with
$$
q_{\lambda} = {\rho^2 \over \lambda^2}\exp [-({\rho^2
\over \lambda^2} + z^2 )]. \eqno(41)
$$
It is clear that, for some fixed, small $C$, for each
$g_{\lambda}$, as $\lambda \rightarrow 0$, one can get a
regular solution to Eqn.(2). Further, the ADM mass along
the sequence remains (more-or-less) constant and no
apparent horizon appears.

\tx Let me now make a (constant) coordinate
transformation on each metric $g_{\lambda}$ replacing
$\rho$ with $\lambda \rho$ and $z$ with $\lambda z$.
This changes (40) to
$$
g'_{\lambda} = \lambda^2 [e^{2Cq'_{\lambda}}(d\rho^2 +
dz^2) + \rho^2d\theta^2], \eqno(42)
 $$
with
$$
q'_{\lambda} = \rho^2 exp[-(\rho^2
 + \lambda^2 z^2 )]. \eqno(43)
$$
Now I conformally transform $g'_{\lambda}$ by dividing
it by $\lambda^2$ to finally get
$$
g''_{\lambda} = e^{2Cq'_{\lambda}}(d\rho^2 +
dz^2) + \rho^2d\theta^2. \eqno(44)
 $$
This combination of coordinate and conformal
transformations cannot change the value of the Yamabe
constant. This means that we can still solve (2) for
each $g'_{\lambda}$. However, the value of the ADM mass
is not a conformal invariant; it picks up a factor
$\lambda^{-1}$. Thus the new mass, instead of
remaining constant, blows up as $\lambda \rightarrow 0$.

\tx When one looks at $q'_{\lambda}$, eqn.(43), it is
clear that the limiting metric is no longer singular.
However, it is still unpleasant, as it is no longer
asymptotically flat. The `z' dependence in $q'$
drops out and the metric becomes cylindrically
symmetric rather than axially symmetric.

\tx The ADM mass is measured with respect to some
`unit' (meters, lightyears, whatever). This combination of
coordinate transformation together with a conformal
transformation  can be regarded as the equivalent of a
change of units, and the value of the ADM mass will
change appropriately. Thus any relationship between the
mass and the Yamabe constant can only be valid in some
very restrictive sense, one has to `fix the units'.

\tx Let me return to the other feature that AHST
observed, the absence of apparent horizons. It does
not matter whether one performs this combination of
constant coordinate and conformal transformations on the
base space or on the physical space. Apparent
horizons at moment-of-time-symmetry are equivalent
to minimal 2-surfaces. Minimal 2-surfaces are
stable under constant scalings. Therefore, if AHST find no
apparent horizons neither will I, even though the
mass blows up for the sequence I consider. This does not
contradict the analysis in Section 4. In that section I
assumed the existence of an `external' region where the
physical geometry was dominated by the conformal
factor. In the sequence of metrics considered here
the support of the base scalar curvature spreads ever
outwards, preventing us from taking advantage of
the increase of M.

\tx It is not even clear what happens to the Yamabe
constant in the limit $\lambda \rightarrow 0$. All I
can say is that it will not pass through zero at
any finite value of $\lambda$. There is a hint in
Wheeler's spherically symmetric toy model that the
Yamabe constant may remain nonzero, even in the limit.

\tx Let us return to (36),but instead of considering a
sequence in which $B^{1/2}a$ approaches $\pi/2$, let us
increase $a$ and simultaneously scale down $B$ so that
$B^{1/2}a$ remains constant (equal to, say, $\pi/4$).
Now, the coefficient of the 1/r term becomes
unboundedly large (it scales with $a$). However, the
value of $\psi$ itself never becomes large, we find
$\psi \leq \sqrt{2}$ holds true no matter how large
$a$ becomes. Hence we never obtain a surface on which
$\psi = 2$. Further, the limit $a \rightarrow \infty$
does not correspond to the appearance of a zero energy
bound state, a solution to (37). Thus, the limit point
that AHST investigate almost certainly does not
correspond to the Yamabe constant going to zero.

\tx This trick, used by AHST, of scaling the sequence so
that the mass remains constant, can also be implemented
for the `regular' sequences discussed in Section 4.
Wheeler had already realised this in 1964. There are
three equivalent choices of scalings:\par
(i)  : Instead of letting the conformal factor $\phi$
go to 1 at infinity, solve (2) with the condition
$\max \phi$ = 1. This was the choice Wheeler made.\par
(ii) : Retain $\phi \rightarrow 1 $ at infinity, but
multiply all the coordinates with the Yamabe constant
$Y$.\par
(iii): Retain $\phi \rightarrow 1 $ at infinity, but
divide all the coordinates by the ADM mass.

\tx With each of these constant scalings, the
appearance of minimal surfaces is uneffected. Now,
however, as one moves along the sequence, as the
minimal surface moves out to infinity, the area of
the horizon shrinks to zero. When it pinches off at
`infinity', it really is at a finite proper distance
from the middle. Further, it does so smoothly, we do
not get a conical singularity, instead we get a
regular closed manifold. As one moves further along
the sequence, into the region of negative Yamabe
constant, I expect that this pinched off point moves
back from infinity and develops into a conical
singularity. Thus we get again an asymptotically flat
manifold, but now one with a naked singularity.

\ha 7 Conclusions \par

\tx What I have been calling conformal superspace,
the space of asymptotically flat, Riemannian,
3-manifolds that can be conformally transformed into
(asymptotically flat) regular manifolds with positive
scalar curvature, is an interesting object. It is the
natural space one is left with if one, in a 3 + 1
analysis of the Einstein equations, factors out
the Hamiltonian as well as the
momentum constraints. If one does not wish to
proceed this far, one can still think of it as an
interesting subset of regular superspace, metrics
on which we expect to find regular, classical solutions of
the constraints.

\tx Conformal superspace, $\tilde{S}$, consists of all
asymptotically flat, smooth, Riemannian metrics with
positive Yamabe constant. The boundary of this space, the
limit points of (smooth) sequences of metrics in
$\tilde{S}$ will, presumably, consist of metrics
which violate one or other of the defining
properties of $\tilde{S}$. In other words we
expect to find metrics which are not
asymptotically flat, metrics which are not smooth,
metrics which are not uniformly elliptic and
metrics for which the Yamabe constant equals zero.

\tx What understanding we have of such issues as the
`size' and `shape' of conformal superspace has basically
been gained by looking closely at the Brill waves. I am
convinced that yet more information can and will be
extracted from them. In my introduction I stated
that they ``offer valuable insight a third of a
century later''. I have every expectation that
this will continue to hold true for another third
of a century.

\hb Acknowledgements \par

\tx I would like to acknowledge my debt to Bobby Beig,
significant parts of this article are just a recasting of
our joint work. I would like to thank Edward Malec for
critically reading the manuscript.

\ha References \par

\bb Abrahams, A., Heiderich, K., Shapiro, S. and
Teukolsky, S. (1992), Phys. Rev. D{\bf 46}, 1452-1463.
\par \bb Beig, R. and \'O Murchadha, N. (1991), Phys. Rev.
Lett. {\bf 66}, 2421-2424. \par
\bb Brill, D. (1959), Ann. Phys. (N.Y) {\bf 7}, 466-467.
\par  \bb Cantor, M. and Brill, D. (1981),
Compositio Mathematica {\bf 43}, 317-330. \par
\bb Eppley, K. (1977), Phys. Rev. D{\bf 16}, 1609-1614.
\par \bb Gilbarg, D. and Trudinger, N. (1983), {\it
Elliptic partial differential equations of second order}
(Springer, Berlin). \par
\bb \'O Murchadha, N. and York, J. W.
(1974), Phys. Rev. D{\bf 10}, 428-436. \par \bb \'O
Murchadha, N. (1987), Class. Quantum Grav. {\bf 4},
1609-1622.   \par
\bb \'O Murchadha, N. (1989), in {\it Proceedings of the
C.M.A.} Vol.{\bf 19}, edited by R. Bartnik (C.M.A.,
A.N.U., Canberra) 137-167. \par
 \bb Schoen, R.
(1984), J. Diff. Geom. {\bf 20}, 479-495.  \par
\bb Wheeler, J. A. (1964), in {\it Relativity, Groups and
Topology}, edited by B. DeWitt and C. DeWitt (Gordon
and Breach, New York) 408-431.

\bye